\documentclass[12pt, a4paper]{article}

\usepackage[utf8]{inputenc}
\usepackage{graphicx} 
\usepackage{hyperref} 
\usepackage{geometry} 
\usepackage{cite}
\usepackage{amsmath,amssymb,amsfonts}
\usepackage{algorithmic}
\usepackage{textcomp}
\usepackage{subfigure}
\usepackage{comment}
\usepackage{url}
\usepackage{etoolbox}
\usepackage{bm}
\usepackage{enumitem}
\usepackage{xcolor}
\title{XRFlux: 
Virtual Reality Benchmark for Edge Caching Systems}
\author{
    Nader Alfares\\
    Pennsylvania State University\\
    \texttt{nna5040@psu.edu} 
    \and
    George Kesidis\\
    Pennsylvania State University\\
    \texttt{gik2@psu.edu}
}
\date{}

\begin{document}

\maketitle

\begin{abstract}
We introduce a Unity based benchmark XRFlux for evaluating Virtual Reality (VR) delivery systems using edge-cloud caching. As VR applications and systems progress, the need to meet strict latency and Quality of Experience (QoE) requirements is increasingly evident. In the context of VR, traditional cloud architectures (e.g., remote AWS S3 for content delivery) often struggle to meet these demands, especially for users of the same application in different locations. With edge computing, resources are brought closer to users in efforts to reduce latency and improve QoEs. However, VR's dynamic nature, with changing fields of view (FoVs) and user synchronization requirements, creates various challenges for edge caching. We address the lack of suitable benchmarks and propose a framework that simulates multiuser VR scenarios while logging users' interaction with objects within their actual and predicted FoVs. The benchmark's activity log can then be played back through an edge cache to assess the resulting QoEs.  This tool fills a gap by supporting research in the optimization of edge caching (and other edge-cloud functions) for VR streaming.
\end{abstract}

\section{Introduction}
Virtual Reality (VR) has seen significant advancements in recent years, driven by the increasing demand for immersive experiences across 
education and entertainment 
applications  \cite{kim2020systematic}. However, VR systems are highly data-intensive, necessitating rapid data delivery to maintain the immersive Quality of user Experience (QoE). Traditional centralized cloud architectures face challenges in meeting the stringent latency requirements of VR applications, particularly when servicing geographically dispersed users \cite{shi2016edge, satyanarayanan2017emergence}. To address these challenges, edge computing has emerged as a promising paradigm, bringing computational resources closer to end-users. By caching content at edge servers, the system can reduce latency and improve QoE of VR streaming \cite{wang2018survey}. Nevertheless, the dynamic nature of VR environments, characterized by rapidly changing fields of view (FoVs) and heterogeneous user demands, poses significant challenges for the design and evaluation of edge-caching strategies.

We present a preliminary VR benchmark, XRFlux, specifically designed for the evaluation of edge-cloud systems supporting VR streaming. We describe a preliminary open-source VR universe with simulated 3D user FoV motion combined with
interactive motion in the virtual environment among multiple users.
User motion is simulated to avoid human subjects (including Institutional Review Board (IRB)) issues.
The benchmark
comprehensively records the objects entering the FoVs (and optionally the predicted FoVs) of all users, annotating these records with, e.g., distances between objects and user avatars (information that may be used to determine the resolution of the object delivered to the user's device). The log of such records is then utilized to create a demand process (XR workload) that serves as the input for the edge cloud infrastructure. Subsequently, the FoVs are re-rendered using content delivered from the (simulated) edge cloud, possibly accounting for potential delays in content delivery and resolution degradation (particularly for objects that are positioned farther within the FoV). The QoE of the users can then be assessed under different edge-cloud designs and configurations.

\section{Motivation and Related Work}
Despite the potential benefits of edge caching for VR, there is a lack of standardized benchmarks that can effectively evaluate the performance of edge caching mechanisms in realistically dynamic settings. Existing benchmarks primarily focus on traditional video streaming \cite{Lottarini2018VBench} or general-purpose data caching scenarios, which do not account for the unique requirements of VR. For instance, traditional video streaming benchmarks are designed to measure performance metrics like playout-buffer underrun rates and average video quality, which are insufficient for capturing the complex interplay of factors affecting VR experiences such as FoV prediction accuracy. Panoramic videos (360 degree videos) also differ from VR since they limit users to rotational movement (i.e., 3 degrees of freedom), where VR allow for rotational and translation movements (i.e., 6 degrees of freedom) \cite{vivo, GROOT20,CaV3}.

Furthermore, conventional data caching benchmarks do not consider the dynamic and interactive nature of VR environments, where user movement and interactions within the virtual space can drastically alter content demand patterns in real-time \cite{tan2021Learning360Video}. This paper aims to bridge these gaps by developing VR-specific edge caching benchmarks that not only capable of measuring the traditional QoE metrics, but also incorporate new metrics tailored to VR, such as visible object presence and resolution in the user FoVs \cite{Wysopal2023} and 
system-performance metrics
 such as latency in rendering 3D environments and the effectiveness of predictive caching strategies in reducing perceived lag and cyber motion sickness.

Thus, one motivation for this work is to develop a comprehensive benchmark tailored specifically for edge caching in VR streaming environments. This benchmark aims to simulate VR/XR scenarios, capturing the complex interaction dynamics among users and their virtual environments, and providing a robust framework for evaluating the performance of various edge-caching and rendering-support strategies under realistic conditions. By doing so, we aim to fill the gap in existing research and provide the community with a valuable tool for advancing the state-of-the-art in VR edge-caching systems.

\section{Simulation Model and User Dynamics}

This section describes the mechanics of an example idealized VR simulation involving ``groupie'' (follower) and ``principal'' (leader) users who navigate through a created 3D universe of ShapeNet objects \cite{ShapeNet2015}. The groupies maintain proximity to a principal, i.e., they are the principal's entourage. The groupies ``belong" to (are part of the entourage of) their nearest principal
based on Euclidean distance. 
All aspects of this example, including user movement model, number of 
users, and number, location and size of the ShapeNet objects, are easily modified by
the researcher.

\subsection{Universe Definition and Initialization}
Consider the universe defined as the cube $[-1,1]^3$. 
Both principals and groupies initially teleport to random locations within the universe.

\subsection{Movement of Principals}
In a ``basic" mobility model,
a principal moves in straight lines at a constant velocity $\bm{v}\in\mathbb{R}^3$. Upon reaching the boundaries of the universe, the principal bounces off at a random angle, altering the direction while keeping the velocity $\bm{v}$ constant. If the position of the principal at time $t$ is $\bm{p}_t$ and it is far from the boundary, the position at a small time increment $d$ seconds later ($d > 0$) is simply given by:
\begin{equation}
\bm{p}_{t+d} = \bm{p}_t + \bm{v} d
\end{equation}

When a principal at position $\bm{p} = (p_1, p_2, p_3)$  reaches a boundary, i.e., causing  $p_k \geq 1$ or $\leq -1$ for some $k$, the direction of $\bm{v}$ changes. For example, if $p_2 = 1$, then $v_2 < 0$; otherwise, $v_k$ changes randomly. Similarly, if $p_1 = -1$, then $v_1 > 0$; otherwise, $v_k$ changes randomly.

\subsection{Movement of Groupies}
Under an example basic mobility model,
the change in position of a groupie, $\bm{g}$, over a small time period $d$ is 
similarly given by
$\bm{g}_{t+d} = \bm{g}_t + \bm{v}_t d$
where $\bm{v}_t$ is the velocity vector of the groupie at time $t$. 
A groupie's velocity vector is more frequently updated as follows:
\begin{equation}
\bm{v}_{t+d} = \bm{v}_t + (\bm{a} + \bm{f}) \cdot d
\end{equation}
Here,
\begin{itemize}
    \item $\bm{a}$ is the attraction vector towards the nearest principal, defined as:
    \begin{equation}
    \bm{a} = \min\left(F_{\text{max}}, \|\bm{p} - \bm{g}\|\right) \cdot \frac{\bm{p} - \bm{g}}{\|\bm{p} - \bm{g}\|}
    \end{equation}
    where $F_{\text{max}}$ is the maximum force magnitude, $\bm{p}$ is the position of the nearest principal, and $\|\cdot\|$ denotes the Euclidean norm.
    \item The ``diffusion" vector 
    $\bm{f}=(f_1,f_2,f_3)$ where the $f_i$ are independent and uniformly random in $[-\delta,\delta]$.
\end{itemize}

To ensure groupies remain within the universe $[-1, 1]^3$, if any component of $\bm{g}_{t+d}$ exceeds the boundary, the groupie can teleport to a random location within the universe or to a feasible location close to its current principal.

\subsection{Field of View (FoV)}
The FoV for a user can be characterized by the following parameters: 
\begin{enumerate} 
    \item The 
    maximum distance from the user at which the FoV is projected. 
    \item The angular width that defines the extent of the FoV. 
\end{enumerate} 
For each user, we also define a {\it predicted} FoV, which serves as a superset of the user's {\it immediate} FoV, anticipating potential changes in viewing direction. 
Even if the user does not change position in the VR environment, their FoV 
may dynamically shift. 
These angular shifts, which simulate rapid head movements (``rubbernecking"), are modeled by periodically adding independent angular perturbations drawn from a normal distribution, ${\sf N}(0, \sigma^2)$, to to the FoV vector. The period of these updates also can be varied
by the researcher, e.g., taken as  
significantly longer than the characteristic time $d$ which governs the user’s  motion in
the VR environment.

\begin{figure*}
    \centering
    \includegraphics[width=\linewidth]{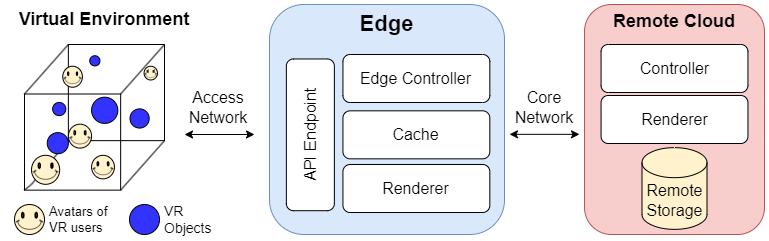}
    \caption{Illustration of the different components of the benchmark.}
    \label{fig:VRBenchSystem}
\end{figure*}

\begin{figure}
    \centering
    \includegraphics[width=0.8\linewidth]{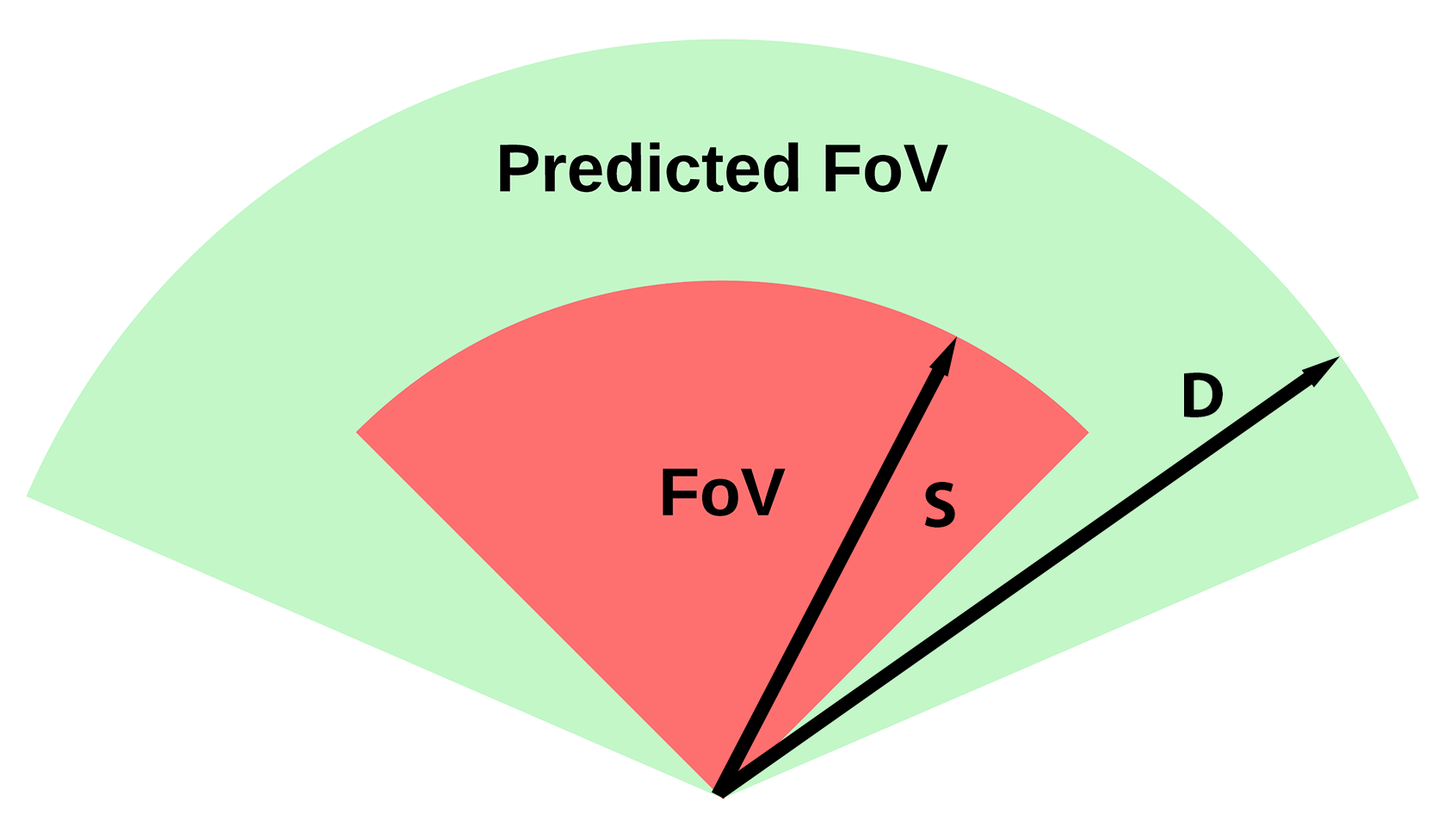}
    \caption{Simple illustrative example of different FoVs per user; where S is depth distance for the immediate FoV, and D is the depth distance for the predicted FoV.}
    \label{fig:FoV_illustration}
\end{figure}

\begin{figure}[ht]
    \centering
    \includegraphics[width=\linewidth]{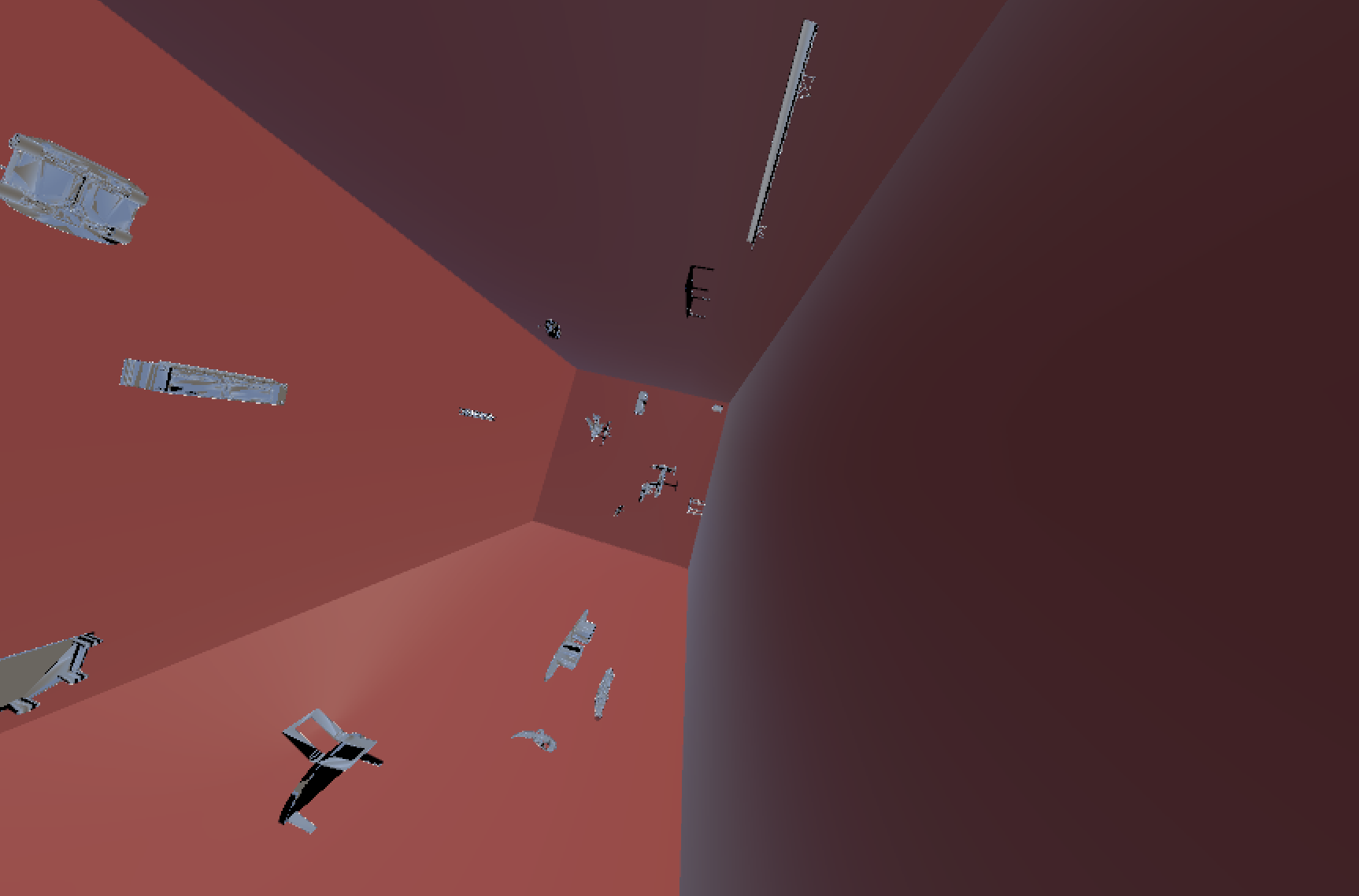}
    \caption{In-game snapshot of the virtual environment from a user's FoV, while in motion. The floating objects in the scene are obtained from ShapeNet \cite{ShapeNet2015}.}
    \label{fig:unitySnapshot}
\end{figure}

\section{Edge Cache for VR}
Edge caching systems aim to minimize latency and reduce the load on remote cloud servers by storing frequently requested content closer to the users. Since storage at edge will be more expensive than remote storage, one of the crucial performance metrics of an edge cache is its cost-effectiveness  while providing high performance (hit rate). In attempt to provide cost-efficacy, various designs of the cache and its policies have been studied and tailored for different applications, e.g., \cite{Memcached,SIEVE24}. For example, Memcached \cite{Memcached} is a high-performance distributed memory object caching system that provides the option to operate a segmented cache, dividing the cache into multiple sections to store different types of data. As an example in the context of VR, segmented caches can be utilized to store high-resolution separately from low resolution objects. In our benchmark, as illustrated in Figure \ref{fig:VRBenchSystem}, VR users access the cache through API endpoints provided by the edge that allows them to request for objects to be fetched,
particularly through a cellular-wireless access network, 
providing low-latency access to cached content. The benchmark also features a remote cloud interface that retrieves content from a remote storage when cache misses occur at the edge.

In the context of VR streaming, objects of varying resolutions can be cached to further optimize performance. This can optionally be done through the use of layered encoding to try to economize on cache memory at the expense of some rendering overhead 
\cite{HKR06,Pak19}.
That is, just caching highest-resolution objects provides the best visual experience but requires significant storage and bandwidth especially when only lower resolution versions of those objects are in demand, potentially leading to delays causing degraded user views. Alternatively, caching the same object at multiple resolutions—low, medium, and high—introduces a trade-off between storage capacity, compute costs, and access delays \cite{BK20,BMS23}. Lower-resolution objects can be retrieved more quickly, reducing latency, while higher-resolution versions can be progressively fetched or generated as needed. Local rendering support at the edge, either using an algorithm \cite{garland1997Quadric} or an AI model \cite{chen2023NeuralGraph}, can dynamically adjust the resolution of objects to meet performance and QoE targets while conserving cache space. This flexibility ensures that the system can balance computational load with access efficiency, offering a seamless VR experience while minimizing resource consumption.
Also, multiple ``virtual" caches can be implemented within the same edge-cache memory space \cite{Stoica16,mcd-os-m4iot}; this concept allows for the partitioning of the edge cache into distinct virtual segments, each allocated to different group of users. So, each ``entourage" can pay for its own virtual cache, thus providing them with a dedicated portion of the overall cache resources. However, to maximize efficiency and reduce redundancy, these virtual caches can ``share" common objects.

\begin{figure}[ht]
    \centering
    \includegraphics[width=\linewidth]{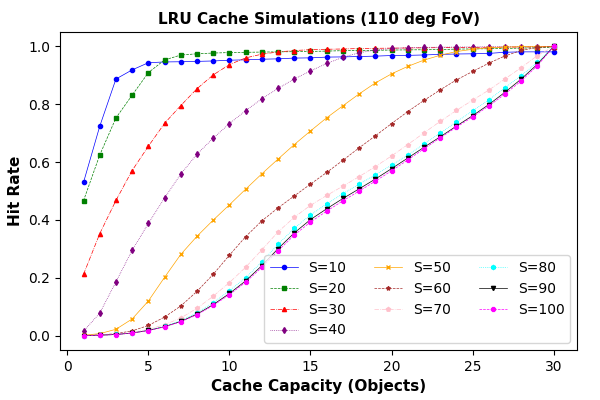}
    \caption{Hit rate of shared cache to all users at different depths (cutoffs) of the FoV at 110\textdegree.}
    \label{fig:cacheHitRateForVariousDist}
\end{figure}

\begin{figure}[ht]
    \centering
    \includegraphics[width=\linewidth]{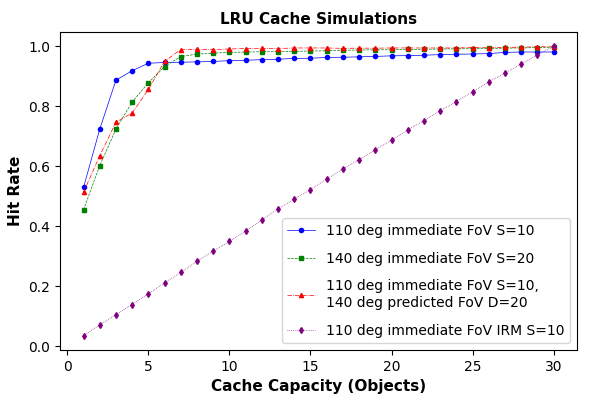}
    \caption{Hit rate of shared cache to all users at three different cache access behaviors. The blue line with circle marker only considers the immediate FoV with angle of 110 degrees and depth $S=10$. The green line with square markers considers the immediate FoV with 140 angle degrees and depth $S=20$. The red line with triangle markers considers both FoVs, immediate and predicted, with objects being pre-fetched at the cache when they are at the predicted FoV. Lastly, the purple line with diamond markers represent the IRM results.
    }
    \label{fig:cacheHitRateFor10_20}
\end{figure}

\begin{figure}[ht]
    \centering
    \includegraphics[width=\linewidth]{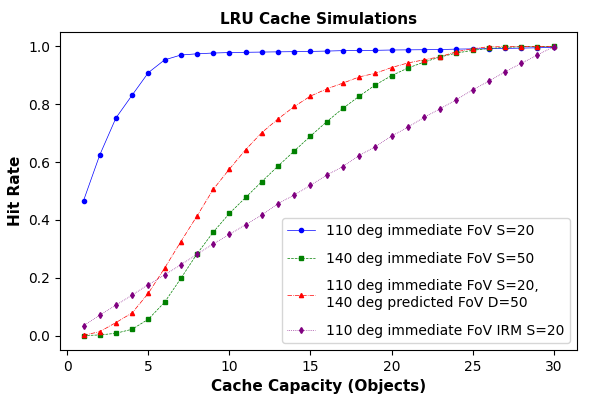}
    \caption{Similar to figure \ref{fig:cacheHitRateFor10_20}, this plot illustrates hit rates of shared cache to all users at three different cache access behaviors. however, in this experiment, we consider depth distances of $S=20$ and $D=50$.}
    \label{fig:cacheHitRateFor20_50}
\end{figure}

\section{Preliminary VR Benchmark Implementation}
Our VR benchmark addresses several limitations of existing approaches to cache performance and QoE evaluation, such as the reliance on the Independent Reference Model (IRM, i.e., independent Poison-process query workloads per object). Additionally, a key motivation for this benchmark is to bypass the need for human-subject experiments, by simulating user movements and interactions, thus offering a controlled and scalable environment when assessing performance and QoE. In our implementation, we consider a simple virtual environment that consists of VR users 
(their avatars) and ShapeNet objects that are placed randomly within the virtual universe. However, our implementation in modeling users motion is portable to different VR applications and environment (an on-going work is to port our implementation to cityscape-like virtual environment, or virtual environments that are dynamically created using Wave Function Collapse (WFC) \cite{wfc}). This allows third-parties (VR application providers) to easily reconfigure our modeling of users motion and tracking of objects in their respective FoVs tailored to fit the VR application scenarios.

\subsection{Implementing the Virtual Environment}
The virtual environment in our benchmark consists of 3D objects from ShapeNet that are retrieved to Unity from an AWS S3 bucket. These objects are used to generate object-request traces, forming the basis of the edge-cache workload.
(Figure \ref{fig:unitySnapshot} shows a snapshot of the virutal environment from a user's prespective). We simulate user motion by modeling them as Unity cameras in the virtual environment. Each user is represented by two synchronized cameras that follow the same motion path and rotation. The first camera captures the user’s immediate FoV, representing the visual space directly in line of sight. The second camera represents the predicted FoV, which anticipates where the user will look next based on movement patterns and behavioral predictions. In one scenario, the predicted FoV’s angle is dynamically adjusted based on the average rate of change observed in the immediate FoV across the three degrees of freedom (3DoF). For instance, if a user exhibits a consistent rightward rotation along the yaw axis (e.g., a 15 degree increase in rotation to the right), the predicted FoV will be adjusted to anticipate this movement by preemptively rotating further to the right. Similarly, changes in pitch or roll will result in corresponding adjustments to the predicted FoV. XRFlux accommodates different FoV prediction methods.
Also note that, in Unity, a user's immediate and predicted FoVs can be 
arranged to produce a combined 360 degree FoV.

Throughout the simulation, we log the visibility of objects within the FoVs of multiple users. Specifically, we track which objects enter and exit each user’s immediate and predicted FoV. By analyzing the logged data, we can simulate user requests for retrieving objects from the cloud.

\subsection{Edge Implementation}
The edge is composed of several key components that work in tandem to ensure efficient content delivery and real-time interaction for users. These components (shown in figure \ref{fig:VRBenchSystem}) include a controller, a cache, a local renderer, and API endpoints that users can invoke when fetching objects.

\begin{itemize}
    \item {\bf The Controller:} It is responsible for fetching objects from a remote server when a cache miss occurs, ensuring that required assets are available for rendering in VR environments. In addition to object retrieval, the controller monitors the edge’s resources, dynamically adjusting their allocation based on demand. This includes resource sizing, which optimizes performance by managing the available computational power and storage at the edge (or navigates a performance/cost trade-off).

    \item {\bf The Cache:} For an initial experiment,
    it operates with a basic Least Recently Used (LRU) replacement policy
    to ensure that commonly used objects are readily available, enhancing responsiveness in VR applications. The cache is designed with modularity in mind, allowing the replacement of the LRU policy with other caching strategies designed by the researcher.

    \item {\bf The Renderer:} The renderer is responsible for generating 3D objects at the edge, offering the ability to dynamically adjust an object's Level of Detail (LoD). This flexibility allows the system to fulfill user requests by rendering lower-resolution versions of objects, when appropriate, instead of fetching higher-resolution assets from the remote server. This approach reduces network load and improves performance, particularly when high-fidelity rendering is not immediately required.

    \item {\bf API Endpoints:} The edge exposes a set of API endpoints that VR users can invoke to fetch objects from the cache.
\end{itemize}

\subsection{Remote Cloud Implementation}
The cloud architecture consists of several components that work together to complement the edge devices and handle more resource-demanding operations. The key components of the remote cloud are:

\begin{itemize} 

\item {\bf Controller:} The cloud controller is responsible for managing and scaling the cloud's computational and storage resources, dynamically adjusting based on real-time demand. The controller also coordinates communication between the cloud and the edge, to transmit objects to the edge cache when a cache miss occur.

\item {\bf Renderer:} Similar to the edge renderer, the cloud-based renderer can process 3D objects and adjust their Level of Detail (LoD) based on the requirements of the VR environment. This allows the cloud to relieve the edge from the computational burden of rendering complex, high-resolution objects.

\item {\bf Storage:} The cloud serves as the central storage repository for VR assets, such as 3D models, textures, and environment data. Leveraging cloud storage services (e.g., AWS S3), the system can store vast amounts of data, including high-resolution versions of all assets that are streamed to the edge upon request.

\end{itemize}

The benchmark is designed to be adaptable for use in third-party VR applications built in Unity. It allows for the integration of users into different virtual environments, with the option to modify the motion model to suit the expected behavior in the target application. For example, in a cityscape environment, the motion model can be adjusted to restrict movement to two axes, reflecting limitations such as navigating along streets.

\begin{figure}[ht]
    \centering
    \includegraphics[width=\linewidth]{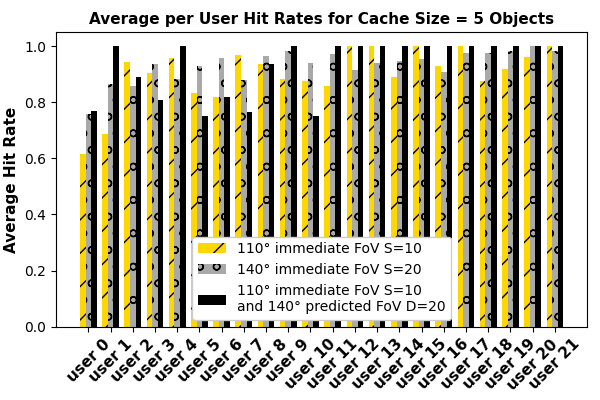}
    \caption{Average per-user hit rates for three cache strategies with a cache size of 5 objects. The bar graph compares the three cache access behaviors for each user.}
    \label{fig:users_hit_rates}
\end{figure}

\begin{figure}
    \centering
    \includegraphics[width=\linewidth]{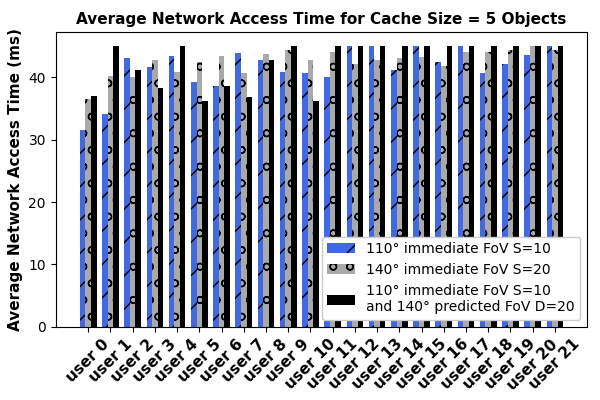}
    \caption{Average per user network access delay with a cache size of 5 objects, with three cache behaviors for each user.}
    \label{fig:users_network_delays}
\end{figure}

\section{Illustrative Experimental Results on Cache Performance}
We simulate a shared cache among all VR users, where users send requests to fetch objects when entering their FoV. On a cache miss, the edge controller retrieves the object from the remote host, updates the cache, and fulfills the user's request. In our experiments, we define three distinct patterns of VR user demand for objects, which are influenced by the state of their FoVs. Specifically for an illustrative example, as shown in figure \ref{fig:FoV_illustration}, we categorize the FoV into two types: the Immediate FoV (with angle of 110 degrees and depth of $S$) and the Predicted FoV (with angle of 140 degrees and depth of $D$). The Predicted FoV is a superset of the Immediate FoV and includes 3D objects that may not necessarily be in the user's direct line of sight but are anticipated to enter their view. These different FoV states impact the patterns of object requests and caching behavior. The three distinct request patterns are:
\begin{itemize}
    \item {\bf Standard FoV Requests:} Users request objects that enter their immediate FoV.
    \item {\bf Extended FoV Requests:}  Users request objects within a larger (predicted) FoV that extends beyond their direct line of sight. This behavior allows users to preemptively fetch objects not yet visible.

    \item {\bf Prefetching Strategy:} The cache prefetches objects that are within the predicted FoV before they are requested, ensuring quicker access to content as it comes into the immediate FoV.
\end{itemize}

Figure \ref{fig:cacheHitRateForVariousDist} illustrates the cache hit rates for the standard FoV requests when objects are within depth (cutoff) $S$ units from the user. As $S$ increases, more ``unnecessary'' objects are stored in the cache; leading to the degradation of cache hit rates.

Figures \ref{fig:cacheHitRateFor10_20} and \ref{fig:cacheHitRateFor20_50} illustrate the cache hit rates for the three different strategies. In each of the figures, we vary the depth (cutoff) the FoVs (both immediate, 110\textdegree depth $S$, and predicted, 140\textdegree depth $D$) and evaluate the shared cache hit rates. We also compare hit rates obtained from the the benchmark to the cache hit rates from a commonly used theoretical Independent Reference Model (IRM) for (standard) requests that are in the immediate FoV; i.e., under the IRM, the queries for each object in the system follows an independent Poisson process
conditioned on the total number of queries for each object are the
same as those logged for simulated user movement.

Using the benchmark, we evaluate the per-user hit rates depicted in Figure \ref{fig:users_hit_rates}, comparing three distinct cache access strategies. This evaluation method allows us to establish a clear correlation between hit rates and the QoE for users in the virtual environment. Specifically, lower hit rates correspond to higher latency as users retrieve objects from the remote server instead of the local cache. This increased latency disrupts the user’s immersive experience in VR applications, where rapid and seamless object rendering is crucial for maintaining engagement and overall QoE. Note that \texttt{user 0} and \texttt{user 1} serve as the principals (leaders) in our experiments. The lower hit rates observed for these principals in the absence of cache prefetching are a result of the user motion design in our setup. Specifically, objects are more likely to first appear in the leader's field of view as part of an entourage.

We generate delays for fetching objects from \cite{Andreev13} for wireless access, and additionally for 
cache misses, from \cite{cloudping} for round-trips to the ``closest'' AWS regions.
A user's QoE subject to such object-fetching delays can then be assessed.
In terms of average object delays per user,
Figure \ref{fig:users_network_delays} compares the three distinct cache-access strategies.




\section{Summary and Future Work}
This paper introduced a preliminary VR/XR streaming benchmark, XRFlux,
to evaluate cache performance at the edge, particularly for multiple mobile users. 
By simulating user movements with dual cameras representing immediate and predicted fields of view (FoVs), the benchmark simulates the dynamic retrieval of 3D objects from the nearby edge cloud (i.e., on a cache hit) or from  the remote cloud (i.e., on a cache miss). The benchmark is flexible and adaptable to third-party VR applications in Unity, allowing, e.g., for customization of user motion models to fit the applications requirements and expected user behaviors. Also note that the dual cameras could be configured as fixed 180\textdegree apart so that the user always demands a 360\textdegree near-FoV. Another extension that could be configured is to consider device-level caching per user. Note that very small form factor (e.g., AR eye glasses) might have very limited memory for caching objects.

Ongoing work is incorporating the Wave Function Collapse (WFC) algorithm \cite{wfc} to dynamically generate 2D and 3D user pathways in another created virtual environment,
and exploring the use of existing open-source cityscapes, where the latter require more computational  resources and more complex constraints on simulated motion. 

We will also support delivery of visible objects at different resolutions depending on their distance to the user. 
Note that in some cases, some portions (view angles) of an object may be much more popularly than others,
so it might be economical to partition objects in addition to differently resolving them.

The benchmark will also account for performance impacts of  emulated edge-transport protocols like QUIC \cite{quic2017}, IT resource overheads and costs.
The benchmark  will accommodate different user QoE measures which depend on such factors
(e.g., object presence and resolution, especially in the near FoV). 
The benchmark will also accommodate alternative edge-caching policies (beyond the LRU strategy considered here, e.g., segmented caches housing multiply resolved objects \cite{HKR06, YangHotOS23, YangNSDI24}) 
and more efficient 3D object rendering-support techniques at the edge to improve performance in resource-constrained environments. Finally, the benchmark will accommodate replaying
workloads through edge-cloud prototypes which are logged from human-subjects
studies, i.e., in which case the movement of user avatars does not need to be simulated.

\section*{GitHub link to open-source software}
The benchmark can be found in a GitHub repository at \href{https://github.com/PSU-Cloud/XRFlux}{\texttt{https://github.com/PSU-Cloud/\\XRFlux}}.

\section*{Acknowledgments}
This work is supported by NSF CNF grant 2212201.
\clearpage


\clearpage
\bibliographystyle{plain}
\bibliography{bibfiles/cache,bibfiles/VR,bibfiles/bari, bibfiles/gk}

\end{document}